\documentclass[sigconf,nonacm]{acmart}

\usepackage{enumitem} 
\usepackage{todonotes}
\usepackage{subcaption}

\AtBeginDocument{%
  \providecommand\BibTeX{{%
    \normalfont B\kern-0.5em{\scshape i\kern-0.25em b}\kern-0.8em\TeX}}}

\copyrightyear{2024}
\acmYear{2024}
\setcopyright{rightsretained}
\acmConference[SIGCSE Virtual 2024]{Proceedings of the 2024 ACM Virtual
Global Computing Education Conference V. 1}{December 5--8, 2024}{Virtual
Event, NC, USA}
\acmBooktitle{Proceedings of the 2024 ACM Virtual Global Computing Education
Conference V. 1 (SIGCSE Virtual 2024), December 5--8, 2024, Virtual Event, NC,
USA}
\acmDOI{10.1145/3649165.3690097}
\acmISBN{979-8-4007-0598-4/24/12}

\makeatletter
\gdef\@copyrightpermission{
  \begin{minipage}{0.3\columnwidth}
   \href{https://creativecommons.org/licenses/by/4.0/}{\includegraphics[width=0.90\textwidth]{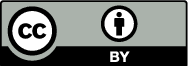}}
  \end{minipage}\hfill
  \begin{minipage}{0.7\columnwidth}
   \href{https://creativecommons.org/licenses/by/4.0/}{This work is licensed under a Creative Commons Attribution International 4.0 License.}
  \end{minipage}
  \vspace{5pt}
}
\makeatother

\begin{document}

\title[``Sometimes You Just Gotta Risk It for the Biscuit'': A Portrait of Student Risk-Taking]{``Sometimes You Just Gotta Risk It for the Biscuit \includegraphics[height=1em]{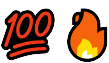}'': \\A Portrait of Student Risk-Taking}


\author{Juho Leinonen}
\affiliation{%
  \institution{Aalto University}
  \city{Espoo}
  \country{Finland}}
\email{juho.2.leinonen@aalto.fi}
\orcid{0000-0001-6829-9449}

\author{Paul Denny}
\affiliation{
    \institution{University of Auckland}
    \city{Auckland}
    \country{New Zealand}
    }
\email{p.denny@auckland.ac.nz}


\begin{abstract}
Understanding how individuals make decisions involving risk is a fundamental aspect of behavioral research. Despite the ubiquity of risk in various aspects of life, limited empirical work has explored student risk-taking behavior in computing education. This study aims to partially replicate prior research on risk-taking behavior in software engineers while focusing on students, shedding light on the factors that affect their risk-taking choices. In our work, students were presented with a hypothetical scenario related to meeting a course project deadline, where they had to choose between a risky option and a safer alternative. We examined several factors that might influence these choices, including the framing of the decision (as a potential gain or loss), students' enjoyment of programming, perceived difficulty of programming, and their performance in the course. Our findings reveal intriguing insights into student risk-taking behavior. Similar to software engineers in prior work, the framing of the decision significantly impacted the choices students made, with the loss framing leading to a higher likelihood for risky choices. Surprisingly, students displayed a greater inclination towards risk-taking compared to their professional counterparts in prior research. Furthermore, we observed that students' performance in the course and their enjoyment of programming had a subtle correlation with their risk-taking tendencies, with better-performing students and those who enjoyed programming being marginally more prone to taking risks. Notably, we did not find statistically significant correlations between perceived difficulty of programming and risk-taking behavior among students.
\end{abstract}

\ccsdesc[500]{Social and professional topics~Computing education}

\keywords{risk-taking behavior, student decision-making, framing effects, academic performance, replication, nudging}



\maketitle

\section{Introduction}

Risk-taking is an inherent aspect of human decision-making, and part of daily life. From minor choices such as trying a new cuisine to more significant professional decisions such as how to best implement a new software feature, many factors contribute to how decisions are made.  More complex decisions often involve carefully weighing up potential benefits against potential losses,  taking into account an individual's appetite for risk.  Quite subtle factors can also play an important role, including how the choices that one is deciding between are presented.  Indeed, the way that options are framed has been shown to significantly influence people's choices~\cite{tversky1981framing}, demonstrating that decision-making is as much about perception as it is about logic.  In educational contexts, if the framing of decisions can impact the choices that students make, we believe this has important implications for educators.

While there exists a multitude of research on risk-taking in other fields~\cite{yates1992risk}, it has received little attention in computing education research. Our work is inspired by the work of Graf-Vlachy in the field of software engineering~\cite{grafvlachy2023risktaking} where the risk-taking behavior of professional software engineers was studied through a survey experiment with 124 participants.  The primary finding was that the way decisions were presented (framed as gains or losses) significantly correlated with software engineers' risk-taking behaviors. However, when adjusting for multiple testing, the study did not find a significant relationship between risk-taking and the Big Five personality traits (see~\cite{roccas2002big}). This prior work highlights that engineers are generally risk-averse, but framing can lead to substantial changes in their level of risk-taking. The results imply that both individual developers and project managers should be cognizant of framing effects when making decisions in software projects.

In our work, we partially replicate Graf-Vlachy's study with students taking an introductory programming course. We examine if, similar to professionals, the framing of the risk-taking scenario correlates with students' decisions. As the original study did not find significant correlations between personality traits and risk-taking, we analyze other attributes. More specifically, we study the correlation between risk-taking and students' performance in the course, their enjoyment of programming, and whether they find programming difficult. We also explore students' rationales for their risk-taking choices. Our research questions for this work are:

\begin{itemize}
    \item RQ1. How does the framing of a risk-taking scenario, the order of choices, the perceived difficulty of programming and programming enjoyment, and course performance correlate with risk-taking?
    \item RQ2. What rationales do students give for their risk-taking choices?
\end{itemize}

\section{Related Work}

Understanding the impact of cognitive biases and individual differences on decision-making has been studied extensively across a variety of domains, including medicine and healthcare \cite{saposnik2016cognitive}, work management \cite{acciarini2021cognitive}, and public policy and governance \cite{decaro2017understanding}.  One robust finding is that the decisions people make can often be heavily influenced by the way that choices are presented to them.  For instance, Tversky and Kahneman describe `framing' effects in which people's choices can be influenced by whether the outcomes are presented as gains or as losses \cite{tversky1981framing}.  

\subsection{Exploring Framing Effects}

The classic example used to study framing effects involves presenting someone with two different options for how a new disease could be managed.  If the options are worded in terms of the number of lives that might be saved (which is a positive frame), people tend to to be more likely to select the option that provides certainty.  In other words, they are less likely to take a `risky' option where the outcome depends on chance, even if there is a possibility that more people would be saved.  Conversely, if the options are presented in terms of the number of lives that might be lost (a negative frame) then people tend to be more likely to select the risky option -- even if the outcomes are statistically equivalent to those in the positively framed scenario.

This phenomenon -- that a choice between a ``sure thing'' and a risky option of equal expected value is affected by option phrasing -- has been widely replicated \cite{fagley1990effect, kuhberger1998influence, dekay2022accelerating}.  Graf-Vlachy~\cite{grafvlachy2023risktaking} explored this in the domain of software engineering, recognizing that risk-taking is inherent in daily software engineering decisions. His study sought to provide empirical data on the subject, which is often overlooked in favor of risk management at the project or organizational level \cite{masso2022terminology}. Professional software developers were given a survey with a software engineering scenario and given two options: a risk-averse and a risk-seeking option.  The framing manipulation involved presenting these options in two distinct ways: as `gains' and as `losses'.

The scenario asked participants to imagine they were working on a software project with a deadline, and that some requirements were wrongly implemented which will result in a missed deadline of 6 weeks.  For participants in the first
condition the options were framed as `gains', i.e., there was a chance of recovering time.  They were presented with the following two options:  (A) If you reduce non-essential features, you will
recover 2 weeks or (B) If you simplify the software architecture, there
is a 1/3 chance that you will recover the full 6
weeks, and there is a 2/3 chance that the simplified
architecture will lead to performance problems and
you will not recover any time at all.

In the second condition, the options were described in
terms of `losses', i.e., the delay with
which they would finish the project, and participants were asked to choose between:  (A) If you reduce non-essential features, you will
finish with a delay of 4 weeks or (B) If you simplify the software architecture, there
is a 1/3 chance that you will finish the project
with no delay at all, and there is a 2/3 chance that
the simplified architecture will lead to performance
problems and you will finish with a delay of 6
weeks.

These subtle differences in phrasing were used to study the framing effect on risk-taking decisions.  No correlations were found between personality traits and risk-taking after multiple testing adjustments, suggesting that individual predispositions might not be as influential as situational framing.  The key finding was the substantial effect of framing: software engineers exhibited higher risk-taking behaviors (i.e., favoring options involving an element of chance) when decisions were framed in terms of losses.  The results imply that both individual developers and project managers should be cognizant of framing effects when making decisions in software projects.

\subsection{Framing Effects in Education}

In educational contexts, framing effects have been found to significantly sway student decision-making processes. For instance, Smith and Smith \cite{smith2009impact} found that the framing of grading systems can impact students' motivation and perceptions. Similarly, Bies-Hernandez \cite{bieshernandez2012effects} highlight that negatively framed academic evaluations can adversely affect students' preferences and performance, and Tansley et al. \cite{tansley2007effects} show that the framing of messages notably affects students' career-related behaviors, particularly when loss-framed. Personal attributes can also impact framing effects, for example Dunegan \cite{dunegan2010gpa} found that students with higher GPAs are more sensitive to framing.  Other personal attributes may impact decision making depending on context.  For example, when decisions relate to performance in a course, they may be heavily influenced by fear of failure which varies significantly between individuals ~\cite{coreia2017fear}.  This body of work on framing and decision making underscores the critical role of framing in shaping student perceptions and their actions.

To the best of our knowledge, there is no research on how framing effects might impact students in computing courses.  However, students are faced with complex decisions as they are studying, and insights into the factors that might influence risk-taking behaviors for students can inform pedagogical strategies.  We seek to replicate the study of Graf-Vlachy, exploring framing effects in the context of a scenario common to computing students -- submitting assignments on time~\cite{denny2021promoting,castro2022experiences,leinonen2021does,leinonen2022comparison}.

\section{Methods}

\subsection{Context}

Our data was collected from a large first-year programming course taught in 2023 at a large public university in New Zealand.  A total of 889 students were enrolled in the course that term, and all were invited to participate in our study for a very small amount of course credit (approximately 0.25\% of the final grade).  

The size of the course is relevant because a sufficiently large number of participants is needed to achieve the statistical power necessary for detecting an effect, if one truly exists, with minimal risk of encountering false positives or negatives.  In the study we are replicating, Graf-Vlachy conduct a detailed power analysis to determine that 124 is the minimum sample size in order to detect a medium effect size with a desired power of 0.8 \cite{grafvlachy2023risktaking}.  The much larger cohort in our study (853 students participated)  allows us to account for additional variability in the data, and detect potentially smaller effect sizes.

The course ran over a total of 12 weeks, and data relevant to this study was collected in weeks 7 and 10.  Given that the scenario we presented students was related to making a decision about a course project and its deadline, this timing ensured that all students had prior experience working on and submitting a project in the first half of the course.  We also used data from the mid-term test, which was conducted in week 6 of the course, to provide a measure of individual student performance.  

At the start of week 7, the mid-point of the course, students were beginning a new module and were asked to reflect on their prior experiences learning to program up to that point.  Responses were collected online through the course learning management system.  The following prompt was shown: ``Now that you have finished your first programming module, read the following statements and reflect on how much you agree with them -- respond on a scale from \emph{Strongly disagree} to \emph{Strongly agree}'', followed by the two statements:

\begin{enumerate}
    \item I find programming enjoyable
    \item I find programming difficult
\end{enumerate}

\subsection{Experiment}

The study had ethics approval from the university's human participants ethics committee. Students were randomly presented one of four scenario options. Two groups had the \textit{gain} framing, and two had the \textit{loss} framing for the scenario. Within each condition, the order of the safe and risky choice was randomly selected, leading to a total of four groups. For each condition, there was a ``safe'' option where the outcome was guaranteed, and a ``risky'' option where the outcome was based on chance, randomly leading to a more positive or negative outcome compared to the ``safe'' option. The expected value was exactly the same for all choices. The number of students in each experiment group is outlined in Table~\ref{tab:students-per-group}.

\begin{table}[ht]
\centering
\caption{Number of students in randomly-assigned experimental groups.}
\label{tab:students-per-group}
\begin{tabular}{lcc}
\hline
Framing/Order & Risky first & Safe first \\
\hline
Gains  & 189  & 192  \\
Losses  & 235  & 237  \\
\hline
\end{tabular}
\end{table}

\subsection{Scenario}

Imagine that you are working on a large software project for a university course with an upcoming deadline on the 10th of the month.  You just realize that you have implemented some of the requirements incorrectly, and you estimate that this will make you miss the deadline by 6 days (i.e. you will submit on the 16th).  Each day lost will reduce the marks that you will receive for the project by 10\% per day (= 60\% in total if you submit on the 16th). You think about how this could be fixed, and you come up with two options. You can only choose one.

\subsubsection{Gain Framing}

Here, the options are framed as ``gains'', or recovering time on the assignment.

\begin{itemize}[leftmargin=2cm]
    \item[\textbf{Option A:}] If you start all over again from scratch, you will \textbf{recover 2 days} (i.e. submit on the 14th).
     \item[\textbf{Option B:}] If you modify the current implementation, there is a \textbf{1/3} chance that you will \textbf{recover the full 6 days} (i.e. submit on the 10th), and there is a \textbf{2/3} chance that modifying the current implementation does not work and you will \textbf{not recover any time at all} (i.e. submit on the 16th).
\end{itemize}

\subsubsection{Loss Framing}

Here, the options are framed as ``losses'', or losing time on the assignment so that a submission will be late.

\begin{itemize}[leftmargin=2cm]
    \item[\textbf{Option A:}] If you start all over again from scratch, you will \textbf{finish with a delay of 4 days} (i.e. submit on the 14th).
    \item[\textbf{Option B:}] If you modify the current implementation, there is a \textbf{1/3} chance that you will \textbf{finish the project with no delay at all} (i.e. submit on the 10th), and there is a \textbf{2/3} chance that modifying the current implementation does not work and you will \textbf{finish with a delay of 6 days} (i.e. submit on the 16th).
\end{itemize}

\noindent After reading the scenario and selecting their choice, students were prompted to provide a rationale by answering the question: \emph{Please explain why you chose that option in a sentence or two}.

\subsection{Variables}

Our study involves a total of six variables. The detailed description of the variables and their use is as follows.

\begin{itemize}
    \item \textbf{Order of the questions}: The order in which the choices were presented to the students was random. This allows examining if the ordering of the choices affects which option is selected. Coded as 0 = risky first, 1 = safe first.
    \item \textbf{Framing of the scenario}: The scenario was framed as gains or losses, similar to the study by Graf-Vlachy. Coded as 0 = losses, 1 = gains.
    \item \textbf{Enjoyment of programming}: Halfway through the course, students were asked to express their level of agreement with the statement \emph{I find programming enjoyable}. Likert-answer between 1 and 5 (1 = ``Strongly disagree'', 2 = ``Disagree'', 3 = ``Neutral'', 4 = ``Agree'', 5 = ``Strongly agree''). 
    \item \textbf{Finding programming difficult}: Halfway through the \linebreak course, students were asked to express their level of agreement with the statement \emph{I find programming difficult}.  Likert-answer between 1 and 5 (1 = ``Strongly disagree'', 2 = ``Disagree'', 3 = ``Neutral'', 4 = ``Agree'', 5 = ``Strongly agree''). 
    \item \textbf{Performance in the course}: We use students' performance in the mid-term examination. Score ranged from 0 to 48 (which was the maximum score available). 
    \item \textbf{Choice of risk}: Whether students chose the risky (uncertain) outcome or the safe (guaranteed) outcome. Coded as 0 = safe, 1 = risky.
\end{itemize}

\begin{figure*}[tp]
\centering
\begin{subfigure}[t]{0.3\linewidth}
    \centering
    \includegraphics[width=\linewidth]{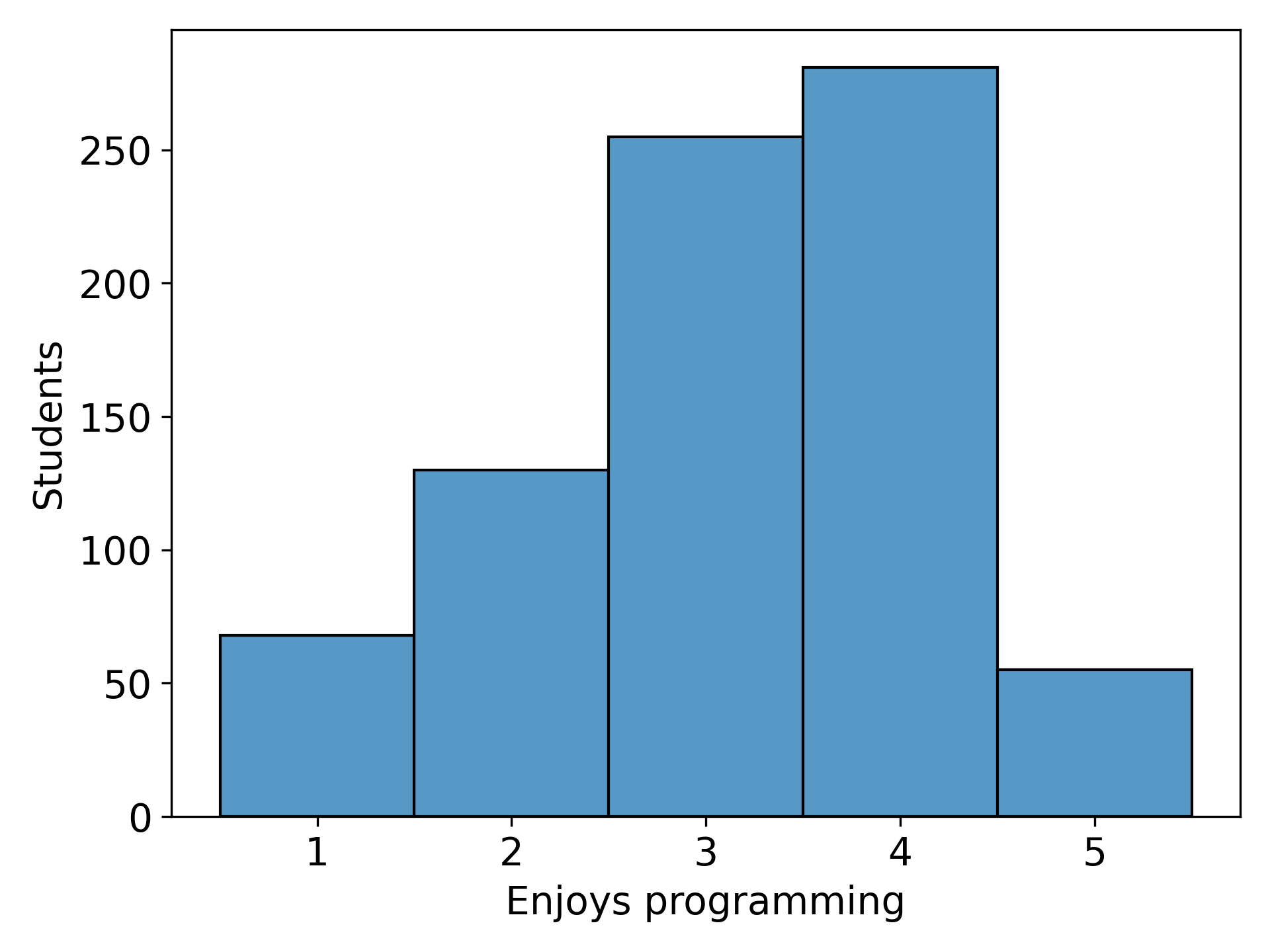}
    \caption{Distribution of student responses to the question of whether they find programming enjoyable.}
\end{subfigure}\hfil
\begin{subfigure}[t]{0.3\linewidth}
    \centering
    \includegraphics[width=\linewidth]{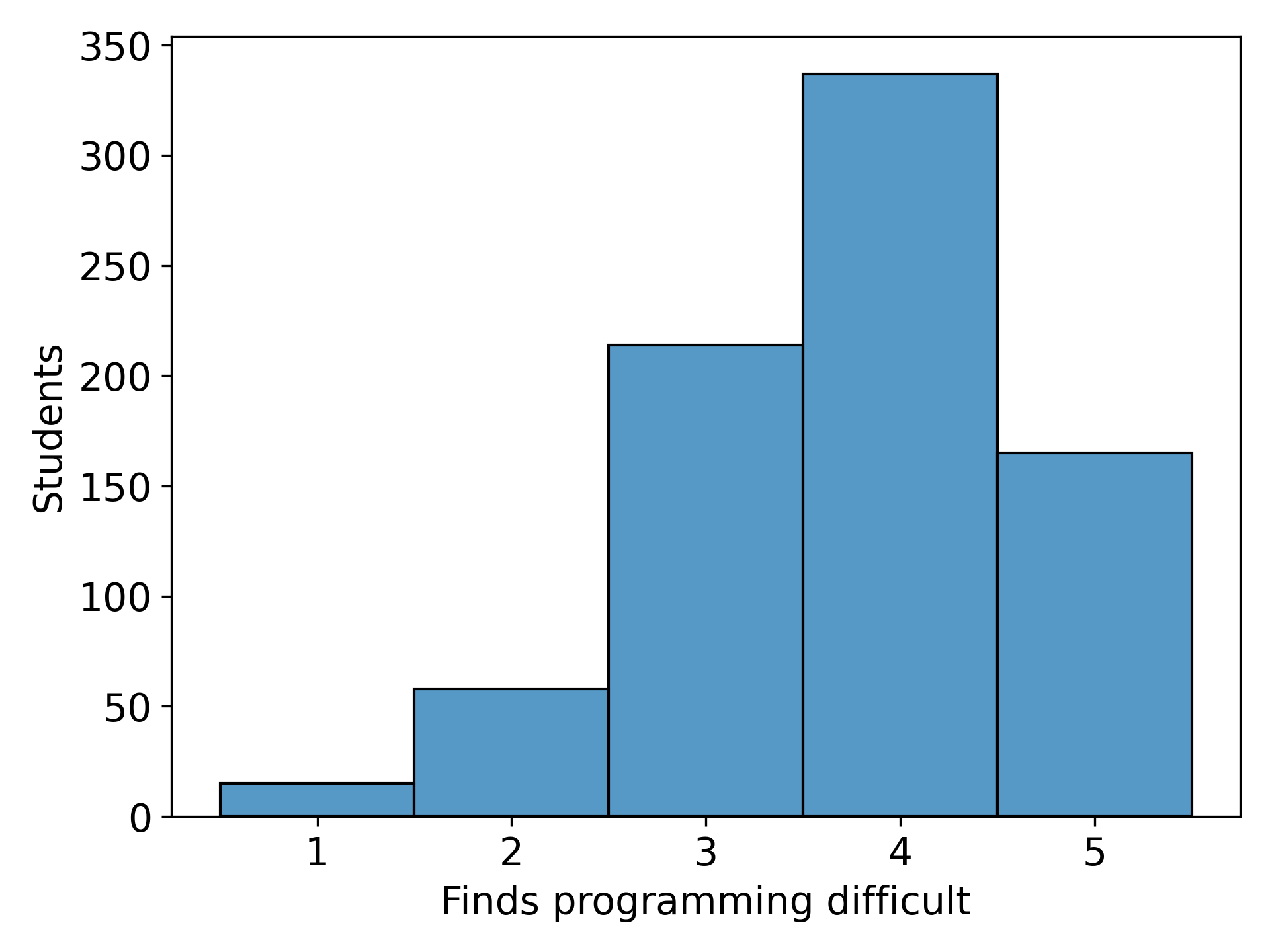}
    \caption[b]{Distribution of student responses to the question of whether they find programming difficult.}
 \end{subfigure}\hfil
 \begin{subfigure}[t]{0.3\linewidth}
    \centering
    \includegraphics[width=\linewidth]{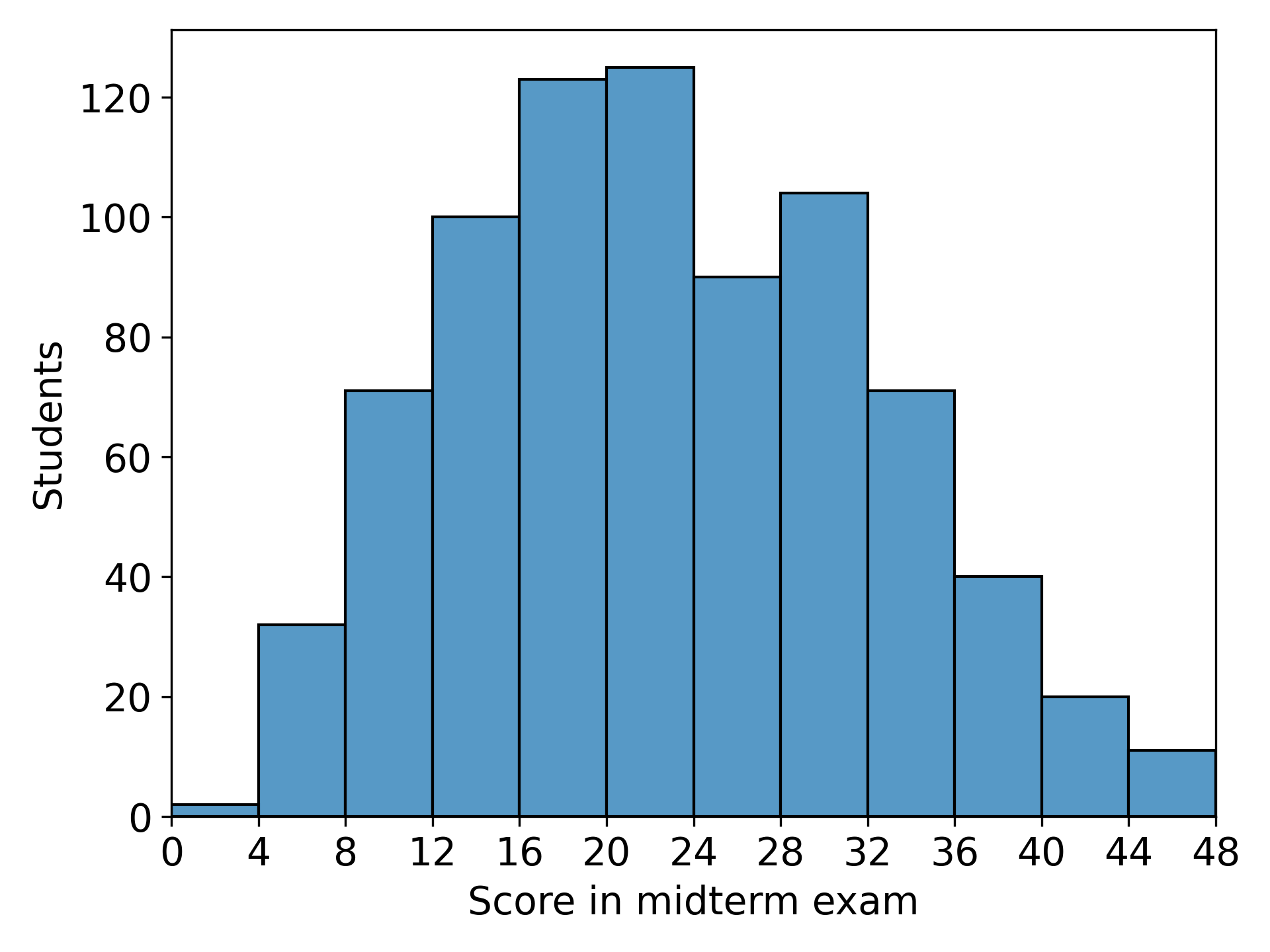}
    \caption[b]{Distribution of student scores in the midterm examination.}
 \end{subfigure}
\caption{Distributions of variables. For (a) and (b), 1 = ``Strongly disagree'', 2 = ``Disagree'', 3 = ``Neutral'', 4 = ``Agree'', and 5 = ``Strongly agree''.}
    \label{fig:descriptive-stats}
\end{figure*}

\subsection{Analysis}

To answer RQ1, \textit{``How does the framing of a risk-taking scenario, the order of choices, the perceived difficulty of programming and programming enjoyment, and course performance correlate with risk-taking?''}, we first conduct a Barnard's exact test\footnote{Using the SciPy~\cite{2020SciPy-NMeth} implementation: \url{https://docs.scipy.org/doc/scipy/reference/generated/scipy.stats.barnard_exact.html}}~\cite{barnard1947significance} to analyze whether there are statistically significant differences between the gain and loss framing (i.e., whether the framing correlates with the choice), and separately if there are differences between the order of choices (i.e., whether the order in which the choices are presented correlates with the choice).

Then, we follow the method of Graf-Vlachy's original study~\cite{grafvlachy2023risktaking}, and run probit regression\footnote{Using the statsmodels~\cite{seabold2010statsmodels} Python package: \url{https://www.statsmodels.org/dev/generated/statsmodels.discrete.discrete_model.Probit.html}} to analyze how the different variables (framing, order, enjoyment, difficulty, performance) correlate with the choice students make in the scenario.

In all statistical analyses, a p-value threshold of 0.05 was set to indicate statistical significance.

To answer RQ2, \textit{``What rationales do students give for their risk-taking choices?''}, we randomly sampled 100 responses to the question ``Please explain why you chose that option in a sentence or two''. Two researchers jointly coded the data for commonly occurring themes, discussing and resolving any disagreements. One response could include multiple themes.

\section{Results}

\subsection{RQ1: How Do Different Factors Correlate With Risk-Taking?}

Table~\ref{tab:choices-by-framing} shows the choices students made with the two different framings of the scenario. For the gain framing, students were almost evenly split between the risky (190) and the safe (191) option. For the loss framing, more students chose the risky option (290 versus 182). The results of a Barnard's exact test confirm that the difference between the framings is statistically significant (statistic = -3.39, p-value = 0.0007).

\begin{table}[ht]
\centering
\caption{The choices students made with the two different framings (gain or loss) of the scenario.}
\label{tab:choices-by-framing}
\begin{tabular}{lcc}
\hline
Framing/Choice & Risky & Safe \\
\hline
Gain  & 190  & 191  \\
Loss  & 290  & 182  \\
\hline
\end{tabular}

\end{table}

Table~\ref{tab:choices-by-order} shows the choices students made with the two different orders of the choices. As can be seen in the table, there seem to be no differences between the group where the risky option was presented first and the group where the safe option was presented first. This is confirmed by the results of a Barnard's exact test  (statistic = 0.61, p-value = 0.55).

\begin{table}[ht]
\centering
\caption{The choices students made with the two different orders of the choices (risky first and safe first).}
\label{tab:choices-by-order}
\begin{tabular}{lcc}
\hline
Order/Choice       & Risky & Safe \\
\hline
Risky-first  & 243  & 181  \\
Safe-first  & 237  & 192  \\
\hline
\end{tabular}

\end{table}

\begin{table*}[ht]
\centering
\caption{Results of the probit regression, showing the coefficient, standard error, z-value, p-value, and 95\% confidence interval. The dependant variable was the choice of risky or safe option, which were coded as 0 (safe) and 1 (risky). The independent variables were framing (0 = losses, 1 = gains), order (0 = risky first, 1 = safe first), whether students enjoy programming, whether students find programming difficult, and performance in the course. The marginal effects show the ``effect size'' of the result, where the dy/dx column shows the change in the probability of making the risky choice with a one unit increase in the variable.}
\label{tab:probit-regression}
\begin{tabular}{l|cccccc|cccccc}
\toprule
& \multicolumn{6}{c}{Probit model} & \multicolumn{6}{c}{Marginal effects} \\
              & Coeff.   & Std. err. & z     & p-value & [0.025 & 0.975] & dy/dx & Std. err. & z & P>|z| & [0.025 & 0.975] \\
\midrule
Framing         & -0.3206 & 0.091   & -3.513 & <0.001 & -0.500 & -0.142 & -0.1218 & 0.034 & -3.600 & <0.001 & -0.188 & -0.055 \\
Order         & -0.0406 & 0.091   & -0.446 & 0.656 & -0.219 & 0.138 & -0.0154 & 0.035 & -0.446 & 0.655 & -0.083 & 0.052 \\
Enjoys        & 0.1172  & 0.049   & 2.384  & 0.017 & 0.021  & 0.213 & 0.0445 & 0.018 & 2.410 & 0.016 & 0.008 & 0.081 \\
Difficult     & 0.0139  & 0.054   & 0.255  & 0.799 & -0.093 & 0.120 & 0.0053 & 0.021 & 0.255 & 0.799 & -0.035 & 0.046 \\
Performance    & 0.0146  & 0.006   & 2.616  & 0.009 & 0.004  & 0.025 & 0.0055 & 0.002 & 2.650 & 0.008 & 0.001 & 0.010 \\
\midrule
const         & -0.4190 & 0.322   & -1.301 & 0.193 & -1.050 & 0.212 & - & - & - & - & - & - \\
\bottomrule
\end{tabular}
\end{table*}

Table~\ref{tab:probit-regression} shows the results of the probit regression. Two factors do not statistically significantly correlate with the choice of risky or safe option: order (p-value = 0.656) and whether the student found programming difficult (p-value = 0.799). However, the other three factors show a statistically significant correlation with the choice: framing (p-value < 0.001), enjoyment of programming (p-value = 0.017), and performance in the course (p-value = 0.009). The marginal effects of the variables on the dependent variable can be seen in the dy/dx column. For example, one point increase in performance (score in the midterm exam) will increase the likelihood of making the risky choice by 0.0055.

Figure~\ref{fig:descriptive-stats} shows the distribution of student responses to the questions on enjoyment and difficulty of programming, and scores in the midterm examination. The distribution of midterm scores is similar to previous iterations of the course.

\subsection{RQ2: What Rationales Do Students Give for Their Risk-Taking Choices?}

Our qualitative analysis of the rationales that students gave for their choices revealed five common themes. We titled the themes ``Risk Tolerance and Aversion'' (46 responses), ``Strategic Calculations'' (37 responses), ``Quality of Work and Professional Work \linebreak Ethics'' (15 responses), ``Mental and Emotional Factors'' (12 \linebreak responses), and ``Optimism versus Pessimism'' (9 responses).

The theme of \textbf{Risk Tolerance and Aversion} emerged as the most commonly mentioned theme, with students weighing the pros and cons of taking risks versus playing it safe. Those who were averse to risk gravitated towards ensuring a known outcome and minimizing any potential negative repercussions. For instance, one student expressed a clear preference for predictability: ``\textit{I would much rather play it safe and have a guaranteed lesser penalty}''. Conversely, those who had a higher risk tolerance recognized a unique opportunity to potentially achieve a perfect score, undeterred by the possibility of incurring a higher penalty, such as the student who declared, ``\textit{Sometimes you just gotta risk it for the biscuit  \includegraphics[height=1em]{emojis.png}}.''

The complexity of decision-making was articulated through the lens of \textbf{Strategic Calculations}, which was the second most common theme. A subset of students approached their decisions analytically, dissecting the probabilities and outcomes at a granular level. One student's response encapsulated this: ``\textit{I would rather have a 100\% chance of having 40\% grade reduction than a 66.667\% chance of a 60\% reduction. Since it means that my max score is 60\% and if I get 50\% I can still pass where as with the 60\% I cannot pass.}'' Their calculations served as the main deciding factor, seemingly offering them a semblance of control in an otherwise uncertain scenario. This also included students who made mistakes in their calculations, incorrectly deducing that one option would have a better expected value than the other: ``\textit{I calculated the probability mathematically. OptionA) Percentage lost = 2 * 10\% = 20\% OptionB) Percentage lost = 1/3 * 0\% + 2/3 * 60\% = 40\% The Percentage Lost is less for option A so I choose option A.}''

The priority for \textbf{Quality of Work and Professional Work Ethics} underscored many students' rationales, influencing them to favor a late submission of higher-quality work over a rushed, on-time submission with potential errors. For many, the potential for high-caliber work, reflective of their skills and effort, outweighed the immediate penalty of delays: ``\textit{Code that works properly is better than broken code in production. I rather something works well than `just works' if not at all.}'' A few students echoed sentiments of completing projects to the best of their ability in line with professional standards or even viewing the scenario as a real-world situation influencing their professional reputation. Their commitment to upholding standards of timeliness and responsibility was captured by one student: ``\textit{I'd rather have a chance to complete the assignment within the deadline as it practices real-world work experiences and by staying calm under immense stress, I can improve my working ethics and efforts.}''

There were also those who included \textbf{Mental and Emotional Factors} in their decision-making process. Stress, anxiety, and emotional well-being were contemplated by students who sought to balance these internal states with the external pressures of their project deadlines. They made decisions based on personal emotional management strategies, with one student lamenting: ``\textit{the amount of time and effort involved in completing a project that is supposed to take 6 full days to complete is not negligible, and to reach the conclusion that you might have to start again from scratch only a couple of days before the deadline is a crushing realization. for this reason i just know i would not have the mental fortitude to scrap all of my work and start again.}''

The theme of \textbf{Optimism versus Pessimism} was also prominent. Optimists relied on the potential promise of perfect project completion within the original time frame, holding onto a hopeful outlook such as, ``\textit{I chose option B as I am simply built different and the odds are always in my favour.}'' Pessimists, however, assumed the likelihood of negative outcomes, expressing doubts in statements like, ``\textit{I've played many luck based games, and I know my luck is usually pretty bad so I would rather option A.}''

\section{Discussion}

We were able to replicate earlier findings~\cite{grafvlachy2023risktaking,fagley1990effect, kuhberger1998influence, dekay2022accelerating} that have suggested the framing of choices correlates with risk-taking behavior. Interestingly, compared to previous results derived from professional software engineers using a similar scenario (i.e., working on a software project), student participants in our study reported being more willing to take risks. Graf-Vlachy ~\cite{grafvlachy2023risktaking} reported 7 out of 63 (11.1\%) participants in the `gain' condition and 19 out of 61 (31.1\%) participants in the `loss' condition chose the risky option.  In contrast, in our case 190 out of 381 (49.9\%) participants in the `gain' condition and 290 out of 472 (61.4\%) participants in the `loss' condition chose the risky option. On the surface, this suggests that students might be more willing to take risks than professional software engineers, however this could also be explained by the fact that the stakes are lower in the context of course projects compared to professional contexts, where there may be impacts on employment or serious consequences for end users of production systems.

Related to the factors that correlate with risk-taking behavior, we were surprised at the result that better performing students were more likely to take risks. Our initial hypothesis was that better performing students would be more similar to professional software engineers, i.e., less inclined to take risks.
Conversely, poor performance may be explained by regular risk-taking involving studying and coursework. However, there are potential explanations for this. First, the study was conducted after the halfway point of the course, and we used the mid-term score as the measure of student performance. Students who did well in the midterm exam had therefore received positive feedback, and thus may have developed higher self-efficacy which may explain their confidence in taking risks to earn a higher mark. Better self-efficacy has been linked with better work-related performance~\cite{stajkovic1998self}, and is one of the strongest predictors of performance in introductory programming as well~\cite{watson2014no,lishinski2016learning}. We also found that students who enjoy programming more were more likely to take risks compared to those who do not. One possible explanation may be that such students are more likely to be excited by the prospect of modifying their current implementation (which was the ``risky'' option in our experiment).

The results of our study have implications for computing instructors. There has been some interest in computing education on ``nudging'' strategies, where  students are gently encouraged towards adopting better learning habits, often through various forms of messaging \cite{brown2021nudging, zamprogno2020nudging, fouh2021nudging, edwards2020proposal,ilves2018supporting,indriasari2023impacting,rogers2021exploring}.  Given the quite large effect we have observed in our work, which simply involved framing options differently, future work should explore how better framing of `nudges' could lead to more impactful results. Another area that could utilize our results is assessment design. Instructors could strategically leverage framing in how assessments are presented. If the goal is to encourage experimentation and creative problem-solving, a loss-framed approach might be effective, as it promotes risk-taking. Conversely, if the focus is on reinforcing foundational skills or ensuring consistent performance, positive framing could help guide students toward safer, more measured choices.

\subsection{Limitations}

Although our aim was to replicate Graf-Vlachy's study~\cite{grafvlachy2023risktaking} as closely as possible, we necessarily made some modifications. \linebreak Firstly, we modified the scenario to involve the deadline for a software project in a university course, to better fit our educational context. This involved, for example, changing the time frame in the scenario from weeks to days which may have had an impact on the results.  Research has found that humans typically take more risks when scenarios involve longer time frames~\cite{bjorkman1984decision}, and thus a change from weeks to days should have reduced risk-taking behavior.  However, we found that the students in our study were more likely to take risks overall compared to the software engineers in Graf-Vlachy's work. Another change that we made is to more clearly explain the consequences of the decision to help students understand the scenario (i.e., we stated precisely how many marks would be lost as penalty for submitting late).  In contrast, the scenario in Graf-Vlachy's work did not concretely specify the consequences for a project delay.

Perhaps the most significant limitation of all relates to the construct validity of using reported decisions on hypothetical scenarios to infer actual risk-taking behavior.  Although there is no reason to believe that students would deliberately provide dishonest answers, if put in the situation where real course outcomes were on the line, they may have acted differently.  This is a difficult limitation to overcome as it would be hard, if not impossible, to conduct a large-scale study of similar nature involving real scenarios.

There may also be other context-dependent factors influencing our results limiting how well they would generalize. For example, the course in which we collected our data is highly competitive, given that it is a required course in an engineering program where entry into preferred second year specializations is determined by GPA.  The competitive nature of the course may affect student risk-taking choices, and a different pattern of responses may be observed in contexts where failing an assignment has lower stakes.

Lastly, there are factors that could affect risk-taking that we did not collect that would have been interesting to study. For example, Graf-Vlachy~\cite{grafvlachy2023risktaking} collected information on participants' programming experience, while we did not. While Graf-Vlachy did not find a statistically significant correlation between programming experience and risk-taking, it is possible that this would be different for students in an introductory programming course. Similarly, it would have been interesting to look at the relationship between gender and risk-taking, as prior research has found that gender correlates with risk-taking behavior~\cite{byrnes1999gender}. However, we did not have this data available to us.

\section{Conclusions}

We conducted an experiment looking at the risk-taking behavior of students in an introductory programming course inspired by an earlier study by Graf-Vlachy~\cite{grafvlachy2023risktaking} that looked at the risk-taking behavior of professional software engineers. We were able to replicate earlier findings suggesting that the framing of choices correlates with risk-taking, and that framing the choices as ``losses'' as opposed to ``gains'' correlates with making riskier choices -- choices where the outcome is based on chance as opposed to a guaranteed outcome. Another interesting finding was that students were in general more inclined to take risks than professional software engineers. Interestingly, students who performed better in the course, and who enjoyed programming more, were somewhat more likely to make risky choices, while whether students found programming difficult or not did not correlate statistically significantly with risk-taking. Lastly, we found that students justified their risk-related choices in different ways, such as based on their risk tolerance, strategic calculations, and emotional factors. Altogether, our study sheds light on risk-taking behavior of students in computing, \linebreak which has been underexplored in previous work.

\begin{acks}
We are grateful for the grants from the Ulla Tuominen Foundation and the Research Council of Finland (Academy Research Fellow grant number 356114) to Juho Leinonen.
\end{acks}

\balance
\bibliographystyle{ACM-Reference-Format}
\bibliography{references}


\end{document}